\documentclass{rspublic}
\usepackage{graphicx}
\usepackage{hyperref}

\begin{document}

\newcommand{\beq}{\begin{equation}}
\newcommand{\eeq}{\end{equation}}
\newcommand{\bqa}{\begin{eqnarray}}
\newcommand{\eqa}{\end{eqnarray}}
\newcommand{\beqa}{\begin{eqnarray}}
\newcommand{\eeqa}{\end{eqnarray}}
\newcommand{\beqan}{\begin{eqnarray*}}
\newcommand{\eeqan}{\end{eqnarray*}}
\newcommand{\nn}{\nonumber}
\renewcommand{\eq}[1]{Eq.~(\ref{#1})}
\newcommand{\erf}[1]{(\ref{#1})}
\newcommand{\dg}{^\dagger}
\newcommand{\bra}[1]{\langle{#1}|}
\newcommand{\ket}[1]{|{#1}\rangle}
\newcommand{\abra}[1]{\langle\underline{#1}|}
\newcommand{\aket}[1]{|\underline{#1}\rangle}
\newcommand{\ip}[1]{\langle{#1}\rangle}
\newcommand{\A}{{\rm A}}
\newcommand{\B}{{\rm B}}
\newcommand{\id}{{{\bf 1}}}
\newcommand{\refcite}[1]{{\cite{#1}}}
\newcommand{\half}{\frac{1}{2}}
\newcommand{\sbin}[2]{\left({}^{#1}_{#2}\right)}
\newcommand{\bin}[2]{\left(\begin{array}{c}
  #1\\#2\end{array}\right)}
\renewcommand{\vec}[1]{{\mathbf{#1}}}

\title[Information erasure]{
{\vspace{-2.3cm}\small\rm \hfill
  {\em Proc. R. Soc. A} (2011) {\bf 467}, $1770–-1778$ \\\vspace{1mm}
  \hfill doi:10.1098/rspa.2010.0577\\\vspace{-3mm}
  \hfill \href{http://dx.doi.org/10.1098/rspa.2010.0577}{http://dx.doi.org/10.1098/rspa.2010.0577}}\\\vspace{1mm}
  \hrule\ \\\vspace{5mm}
 \Large Information erasure without an energy cost}

\author[J.A. Vaccaro and S.M. Barnett]{Joan A. Vaccaro$^{1,2}$ and Stephen M. Barnett$^2$}
\affiliation{$^{1}$Centre for Quantum Dynamics, Griffith University,
Brisbane,\\
Queensland 4111 Australia\\
$^{2}$Department of Physics, SUPA, University of Strathclyde,\\ Glasgow G4
ONG, UK}

\maketitle

\begin{abstract}{information erasure, thermodynamics, canonical ensemble, spin
system} Landauer argued that the process of erasing the information stored in
a memory device incurs an energy cost in the form of a minimum amount of
mechanical work. We find, however, that this energy cost can be reduced to
zero by paying a cost in angular momentum or any other conserved quantity.
Erasing the memory of Maxwell's demon in this way implies that work can be
extracted from a single thermal reservoir at a cost of angular momentum and
an increase in total entropy.  The implications of this for the second law of
thermodynamics are assessed.
\end{abstract}
\hrule

\section{Introduction}

The idea of a link between information and thermodynamics can be traced back
to Maxwell's famous Demon, a supposed microscopic intelligent being, the
actions of which might present a challenge to the second law of
thermodynamics (Maxwell 1871, Leff \& Rex 1990, 2003).  This idea was made
quantitative by Szilard (1929) who showed, by means of a simple one-molecule
gas, that information acquisition, for example by a Maxwell Demon, is
necessarily accompanied by an entropy increase of not less than $k\ln(2)$,
where $k$ is Boltzmann's constant. A closely related phenomenon is the
demonstration, due to Landauer (1961), that erasing an unknown bit of
information requires heat to be dissipated, amounting to not less than
$kT\ln(2)$, where $T$ is the temperature of a thermal reservoir (Plenio \&
Vitelli 2001, Maruyama \emph{et al.} 2010).

Changing the temperature of the reservoir changes the energy cost of erasure
and so it might be argued that it is the fixed entropy of $k\ln(2)$ that is
the fundamental cost of erasing one bit.  One might then be tempted to argue,
erroneously, that energy is not a consideration at all.  However the use of
thermodynamic reservoirs at some temperature is explicit in Szilard's and
Landauer's work. Also, the term $k\ln(2)$ represents {\it thermodynamic
entropy} where the dimension of $k$ is energy per temperature. An energy cost
is therefore inescapable in these analyses except in the extreme case of zero
temperature.

We show here, however, that it is possible to avoid an energy cost,
irrespective of the temperature (Vaccaro \& Barnett 2009).  All that is
required is a different kind of reservoir such as one based on angular
momentum rather than on energy. We begin by recasting Landauer's erasure in
terms of quantum information theory. We then introduce a spin reservoir and
show how it can be used to erase information at a cost in terms of angular
momentum and without a cost in energy and conclude with a discussion.

\section{Erasure using a thermal reservoir}

In quantum information theory, the fundamental component is not the bit, but
rather the qubit (Nielsen \& Chuang 2000, Barnett 2009).  A qubit may be any
quantum system with two distinct states, which we label $|0\rangle$ and
$|1\rangle$. The distinguishing feature of a qubit, of course, is that it can
be prepared in any superposition of these two states. It is instructive to
recast Landauer's erasure principle for a qubit using what we shall refer to
as \emph{Model A} of qubit erasure.  Let us suppose that the qubit is in an
initially unknown state and that we wish to reset it by forcing it into the
state $|0\rangle$. Let the two states $|0\rangle$ and $|1\rangle$ be
initially degenerate, with energy $0$.  We can erase the qubit state by
placing it in contact with a reservoir at temperature $T$ and then inducing
an energy splitting between the qubit states so that $|0\rangle$ has energy
$0$, but the state $|1\rangle$ has energy $E$.  The splitting is induced
adiabatically, that is, sufficiently slowly that the qubit remains in thermal
equilibrium with the thermal reservoir. The state of the qubit when the
energy splitting is $E$ is governed by the Boltzmann (or maximum entropy)
distribution with density operator
\beq
   \rho = \frac{\ket{0}\bra{0}+e^{-E/kT}\ket{1}\bra{1}}{1+e^{-E/kT}}\
   .
\eeq
The work required to increase the splitting from $E$ to $E+dE$ while in
contact with the reservoir is given the probability of occupation of the
state $\ket{1}$ multiplied by $dE$, that is
$dW=e^{-E/kT}(1+e^{-E/kT})^{-1}dE$. The total work in increasing the
splitting from zero to infinity is $W=\int dW=kT\ln2$ and the state of the
qubit is $\ket{0}$ as expected. The qubit is removed from the reservoir and
the energy degeneracy then restored. The erasure here is driven by maximising
the entropy subject to conservation of energy as the energy gap of the states
of the memory qubit grows.

\section{Erasure using a spin reservoir}

Consider a second model of qubit erasure, \emph{Model B}, in which the qubit
logic states $\ket{0}$ and $\ket{1}$ are associated with different
eigenvalues of a conserved observable other than energy. For definiteness, we
take this observable to be the $z$ component of angular momentum and the
qubit to be a spin-$\half$ particle. Let the reservoir be constructed of
similar particles.  In Model A we were able to increase the energy splitting
between the logical states $\ket{0}$ and $\ket{1}$ of the memory qubit by
some external means, and thermalisation with the reservoir involved the
exchange of energy. The erasure of the memory qubit in Model B, however,
proceeds via the exchange of discrete quanta of angular momentum with the new
reservoir. Conservation of angular momentum requires the memory qubit and
reservoir to have compatible angular momentum splittings. We keep the
reservoir splittings fixed and change the effective angular momentum
splittings of the memory system. For this we use a supply of ancilla
spin-$\half$ particles which are initially in the logical zero state, and a
specific controlled operation that ensures the combined memory and ancilla
system is in a superposition of two states which are separated in angular
momentum by a given multiple of $\hbar$. To emphasise their physical context,
we shall hereafter refer to the qubits in Model B as spins.

\subsection{Construction of the spin reservoir}

Let the reservoir consist of $N$ spin-$\half$ particles whose spatial degrees
of freedom are in thermal equilibrium with a much larger heat bath at
temperature $T$. In contrast, the internal spin degrees of freedom are in an
\emph{independent} equilibrium state. The decoupling between the spatial and
spin degrees of freedom may be ensured by requiring the internal spin states
of each spin to be degenerate in energy. The spatial degrees of freedom are
described by a probability distribution which depends on the temperature of
the heat bath. The temperature and spatial degrees of freedom requires no
further consideration.

We describe the internal spin degree of freedom of each spin in terms of the
eigenstates of the $z$ component of spin angular momentum, $S_z$, which we
label in the logical basis $\ket{0}$ and $\ket{1}$. Here $\ket{i}$ represents
an eigenstate of $S_z$ with eigenvalue $(i-\half)\hbar$. The internal states
of the collection of spins in the reservoir is then given by
$\bigotimes_{i=1}^N\ket{x_i}$ where $x_i=0,1$. Consider the set of these
states for which $\sum_{i=1}^N x_i=n$ is an integer in the range $0\le n\le
N$. Each of these states has a $J_z$ eigenvalue of $(n-N/2)\hbar$ and there
are $\sbin{N}{n}$ such states, where $J_z$ is the $z$ component of the
combined spin angular momentum of the whole reservoir. We label elements of
this set as the collective state $\ket{n,\nu}_{r}$ where
$\nu=1,2,\cdots,\sbin{N}{n}$ uniquely labels each state in the set. Thus a
convenient basis for the state space of the spin degrees of freedom of the
reservoir is given by $\{\ket{n,\nu}_{r}\,:\ n=0,\ldots,N;
\nu=1,2,\ldots,\sbin{N}{n}\}$.

We imagine that the internal spin state of the reservoir is generated and
maintained by interacting through the exchange of spin angular momentum with
a much larger ``spin bath''. The spin bath consists of $M$ spin-$\half$
particles with a basis set $\{ \ket{m,\mu}_{b}\,:\ m=0,\ldots,M;
\mu=1,2,\ldots,\sbin{M}{m}\}$ .  Here the collective state of the bath spins
$\ket{m,\mu}_{b}$ represents the $\mu$-th eigenstate of $J_z$ with eigenvalue
$(m-M/2)\hbar$ where $\mu=1,2,\ldots\sbin{M}{m}$ and is defined analogously
to that of the reservoir state $\ket{\cdot,\cdot}_{r}$ . The spin bath
maintains the reservoir in a state such that the average $z$ component of
spin of the reservoir is given by $\ip{J_{z}}=(\alpha-\half)N\hbar$ for
$0\le\alpha\le 1$.

\subsection{Information Erasure}

We treat the reservoir as a canonical ensemble, but instead of energy being
exchanged between the reservoir and spin bath, as is the usual case, the
systems randomly exchange $z$ component of spin according to the mapping
$\ket{n,\nu}_r\ket{m,\mu}_b\leftrightarrow\ket{n\pm1,\nu^\prime}_r\ket{m\mp1,\mu^\prime}_b$
where $\nu^\prime=1,2,\ldots\sbin{N}{n\pm1}$ and
$\mu^\prime=1,2,\ldots\sbin{M}{m\mp1}$. At equilibrium, the probabilities
$P_{n,\nu}$ of finding the reservoir in the state $\ket{n,\nu}_r$ is given by
maximising the {\it information-theoretic entropy} $(-\sum_{n,\nu} P_{n,\nu}
\ln P_{n,\nu})$ with respect to $P_{n,\nu}$ subject to the constraints
$\sum_{n,\nu}n P_{n,\nu}=\alpha N$ and $\sum_{n,\nu} P_{n,\nu}=1$. The
equilibrium probability distribution is found to be
\beq
   P_{n,\nu}=\frac{e^{-n\gamma\hbar}}{(1+e^{-\gamma\hbar})^N}
   \label{P_nm}
\eeq
where $\gamma  = \ln[(1-\alpha)/\alpha]/\hbar$.

Now let us suppose another spin-$\half$ particle is our memory qubit. It
begins in the maximally mixed state $(\ket{0}\bra{0}+\ket{1}\bra{1})/2$ and
we wish to erase its memory and leave it in the logical zero state
$\ket{0}\bra{0}$. Let there be a large collection of ancillary spin-$\half$
particles in the $\ket{0}\bra{0}$ state for our use. The first stage entails
putting the memory spin in spin-exchange contact with the reservoir, letting
the combined reservoir-memory spin system come to equilibrium, and then
separating the memory spin from the reservoir. At this point the state of the
memory spin is
\beq
   p_0\ket{0}\bra{0}+p_1\ket{1}\bra{1}
   \label{state of memory spin 1}
\eeq
with $p_1=e^{-\gamma\hbar}/(1+e^{-\gamma\hbar})=1-p_0$.  We have assumed that the value of
$\ip{J_{z}}=(\alpha-\half)N\hbar$ is maintained by the spin bath and so the value of
$\gamma$ remains fixed despite the contact with the memory spin.

In model A the energy splitting of the memory qubit is slowly increased while
the qubit is in equilibrium with a thermal reservoir. We want the same
principle to operate for Model B but with angular momentum in place of
energy.  In Model B, the $z$ component of angular momentum of the two states
that represent the memory are separated in value by $\hbar$. This separation
can be increased by performing a controlled-not (CNOT) operation (Nielsen \&
Chuang 2000, Barnett 2009) on the memory and an ancilla spin as follows. The
memory spin acts as the control, and the ancilla spin, which is initially in
the state $\ket{0}\bra{0}$, acts as the target. The relevant properties of
the CNOT operator $U$ are given by
\beqa
  U\ket{0j}= \ket{0j}\ ,\qquad
  U\ket{1j}= \ket{1k}\ ,
  \label{mapping CNOT}
\eeqa
where $j$ is $0$ or $1$, $k=1-j$ and, for convenience, $\ket{xy}$ represents
the tensor product of the state $\ket{x}$ of the memory spin and $\ket{y}$ of
the ancilla spin. After the CNOT operation the combined memory-ancilla state
is $p_0\ket{00}\bra{00}+p_1\ket{11}\bra{11}$.  The average angular momentum
cost of this operation is $\hbar p_1=\hbar
e^{-\gamma\hbar}/(1+e^{-\gamma\hbar})$. The combined memory-ancilla system is
placed in random spin exchange contact with the reservoir while the
reservoir-bath is undergoing random spin exchange as before. The spin
exchange between the reservoir and the memory-ancilla system is constructed
to leave all states unchanged except for the following mapping
\beq
  \ket{2,1}_r\ket{00}\leftrightarrow\ket{0,1}_r\ket{11}
  \label{mapping0011}
\eeq
where $\ket{n,\nu}_r\ket{ij}$ represents the reservoir collective state
$\ket{n,\nu}_r$ and memory-ancilla system state $\ket{ij}$. The random spin
exchange continues for a sufficient time for the reservoir and memory-ancilla
system to equilibrate. The state is then given by
\beq
   p_0\ket{00}\bra{00}+p_1\ket{11}\bra{11}
\eeq
where now $p_1=e^{-2\gamma\hbar}/(1+e^{-2\gamma\hbar})=1-p_0$ from \eq{P_nm}.
This completes the first cycle.

Another ancilla spin is added and a CNOT operation is performed as before to
yield the state $p_0\ket{000}\bra{000}+p_1\ket{111}\bra{111}$ with a spin
cost of $\hbar e^{-2\gamma\hbar}/(1+e^{-2\gamma\hbar})$. The combined
memory-ancilla system put in spin-exchange contact with the reservoir with
the mapping
\beq
  \ket{3,1}_r\ket{000}\leftrightarrow\ket{0,1}_r\ket{111}\ .
  \label{mapping000111}
\eeq
This process is repeated.  After $n$ cycles, the memory-ancilla spins are in
the logical zero state and the logical 1 state with probabilities $p_0$ and
$p_1$ where $p_0=e^{-n\gamma\hbar}/(1+e^{-n\gamma\hbar})=1-p_0$ according to
\eq{P_nm}. In the limit of many repetitions and large $N$, the memory-ancilla
system approaches a pure state where each spin is in the logical zero state.
The total spin cost of the whole process is
\beq
    \Delta J_z=\sum_{n=1}^\infty
    \hbar \frac{e^{-n\gamma}}{1+e^{-n\gamma}}\ .
\eeq
This sum is bounded by
\beq
   \gamma^{-1}\ln(1+e^{-\gamma\hbar}) < \Delta J_z <
   \gamma^{-1}\ln(2)\ .
\eeq
If we include the spin of the initial state, then the cost is
\beqa
    \Delta J_z^\prime=\sum_{n=0}^\infty
    \hbar\frac{e^{-n\gamma\hbar}}{1+e^{-n\gamma\hbar}}=\Delta J_z+\half\hbar\
    ,\\
   \gamma^{-1}\ln(2) < \Delta J_z^\prime <
   \gamma^{-1}\ln(1+e^{\gamma\hbar})\ .
   \label{spincost}
\eeqa
Clearly the costs associated with erasure depend on the physical qubit, and need not
include an energy term, in contradistinction to the suggestion of Landauer and many
others.

\subsection{Absence of an energy cost}

There are two important points to be made about the absence of an energy cost
in Model B. First, the cost of erasure is in terms of the quantity defining
the logic states, which is spin angular momentum. This cost arises because
the CNOT operation in \eq{mapping CNOT} does not conserve this angular
momentum. In contrast, the conservation of energy is trivially satisfied due
to the energy degeneracy.

For the second point, note that the energy-free cost of Model B rests
primarily on the decoupling of the internal spin states from any surrounding
thermal reservoir. The decoupling is assured by the energy degeneracy of the
spin states $\ket{0}$ and $\ket{1}$.  However, in any given physical
implementation, there will be limits to the accuracy with which the energy
degeneracy condition could be met. For example, there may be weak residual
magnetic fields in the vicinity of the spins that would lift their energy
degeneracy and produce an energy cost, $\epsilon$, associated with the CNOT
operation. Nevertheless it would still be possible, in principle, to incur an
energy cost much less than $kT\ln(2)$ per bit erased, where $T$ is the
temperature of the thermal reservoir, if the erasure protocol could be
completed in a sufficiently short time. The time scale on which non-ideal
effects become significant is determined by the coupling strengths associated
with energy exchange with the surrounding reservoir, which can be small.
Provided that the erasure protocol can be completed in a time much shorter
than this coupling time, the internal reservoir state would be essentially
independent of the temperature of the thermal reservoir. The average energy
cost of the CNOT operations would then be in proportion to the average spin
cost according to
\beq
   \Delta E\approx\frac{\epsilon}{\hbar}\Delta J_z< \frac{\epsilon}{\hbar}\frac{\ln(2)}{\gamma}\ ,
\eeq
neglecting the initial energy of the memory spin,  which is independent of
the temperature $T$.  {\it The key issue is to perform the erasure protocol
before the system reaches thermal equilibrium.}  It suffices to retain
equilibrium only in the spin degrees of freedom.  The lower bound of
$kT\ln(2)$ on the energy cost, which is associated with thermal equilibrium,
is thereby avoided.

\section{Discussion}

These results open up a range of topics for investigation. For example, the
operation of Carnot ``heat'' engines operating with angular momentum
reservoirs and generating angular momentum effort (or some other resource)
instead of mechanical work. Another possibility is the use of a {\em
combination of different types of reservoir}. For example, a Maxwell's demon
can operate on a single thermal reservoir to extract work from the reservoir.
However there is an associated unmitigated cost in that the memory of the
demon has to be erased to complete a cycle of operation. Bennett's argument
(1982) is to use Landauer's erasure principle to do this, but the extracted
work is more than balanced by the cost of erasure. Given the results above,
we now know that the memory of the demon can be erased using an entirely
different reservoir at no cost in energy.  The cost instead could be in terms
of spin angular momentum as illustrated in Fig.~1.  In this scheme the
demon's memory is assumed to be represented by energy-degenerate spin states.
In step (a) of the figure the demon has no memory and the thermal reservoir
is in equilibrium.  The demon traps the fastest moving molecules in the right
partition of the reservoir in step (b). This allows the demon to extract work
from the reservoir using a heat engine.  In the last step (c) the demon's
memory is erased using a spin reservoir at a cost of spin angular momentum
and the reservoir is allowed to return to equilibrium.  The reservoir and the
demon's memory have completed a full cycle in terms of memory storage,
however, the reservoir now has less heat and correspondingly less
thermodynamic entropy.  The information-theoretic entropy of the spin
reservoir is nevertheless higher than at the beginning of the cycle due to
the erasure of the demon's memory. The scheme represents a cyclical process
of {\it extracting work from a single heat reservoir} at a cost of another
resource (here angular momentum) and a higher overall information-theoretic
entropy.

\begin{figure}
\centering
\includegraphics[width=85mm]{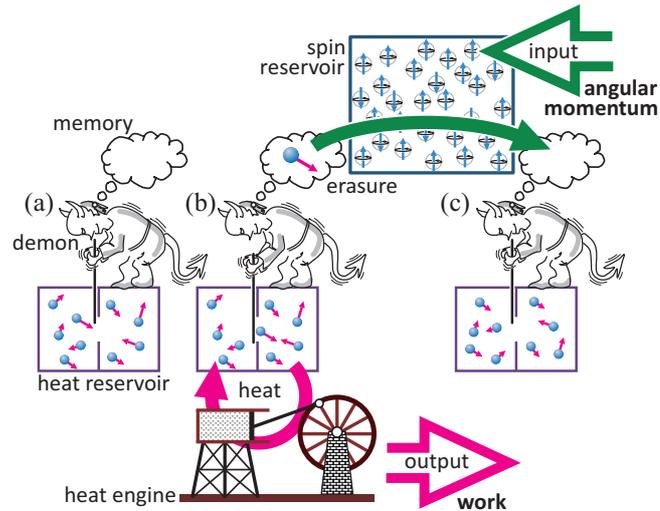}
\caption{\label{fig}  Maxwell's demon extracting work from a single heat
reservoir at a cost of spin angular momentum. In step (a) the demon has no
memory and the gas in the heat reservoir is in thermal equilibrium. Next, in
step (b), the demon performs measurements of the speeds of the molecules and
partitions the reservoir in two, trapping the fastest moving molecules in the
partition on the right side and uses a heat engine operating between the two
partitions to extract work.  Finally in step (c) the demon's memory is erased
using a spin reservoir and the two partitions are allowed to return to
equilibrium. (Online version in colour.)} 
\end{figure}

Although this may appear to be a contentious result, it should not
necessarily be regarded as contradicting various historical statements of the
second law of thermodynamics within their intended contexts.  For example,
consider Kelvin's dictum ``It is impossible, by means of inanimate material
agency, to derive mechanical effect from any portion of matter by cooling it
below the temperature of the coldest part of the surrounding objects''
(Kelvin 1882, p179). The presence of the demon in our analysis, which in
principle could be an automated machine, is not of any significance here.
Rather, Kelvin's discussions are exclusively within the context of heat and
thermal reservoirs which were of overriding importance at the time of his
work and, quite naturally, he did not allow for a broader class of reservoirs
of the kind considered here. Our analysis therefore lies outside of Kelvin's
considerations and within a more general context. For example, our results do
not appear contentious at all for an analogous, but broader, statement of the
second law as: It is impossible to derive mechanical effect from any portion
of matter through a reduction in the information-theoretic entropy of the
system as a whole. The foregoing discussion illustrates the potential impact
of a no-energy-cost erasure protocol.

A quite different approach to this problem has been considered recently by
Sagawa and Ueda (2009).  They explored the possibility of reducing the
information erasure cost at the expense of incurring an additional cost in
the measurement process that initially stores the information. Their main
result is that the erasure process can incur a cost less than $kTI$ where $I$
is the mutual information shared between the memory and the measured system.
The reduction, however, is more than compensated by the cost of the prior
measurement which ensures that the total cost of measurement and erasure is
bounded below by $kTI$.  This is in accord with Landauer's principle for the
ideal case where $I$ is also the entropy of the memory device.  In contrast,
our analysis is in the conventional framework where the measurement process
has a zero energy cost.  The total energy cost of measurement and erasure in
our case is not bounded below by Landauer's energy bound. Rather we have
shown that the cost of erasure can be in terms of another conserved quantity
such as angular momentum.

We wish to emphasise that we have focused here on the principle of
information erasure without an energy cost of the form $kT\ln(2)$.  We
acknowledge that the physical effort to realise these protocols would be
significant especially for large scale reservoirs. But this does not lessen
the conceptual significance of our results. On the contrary, by giving an
explicit example, we have demonstrated that \emph{physical laws do not forbid
information erasure with a zero energy cost on principle}. Practical
limitations, like the accuracy at which the degeneracy condition can be met,
might well indicate that in some given physical implementation there is an
unavoidable nonzero energy cost associated with erasure. But there is no
fundamental reason to suppose that this cost will necessarily be as large as
$kT\ln(2)$. Moreover, the actual energy cost of erasure in such cases would
depend on the particular physical implementation being considered. What is
important here is the lower bound of this cost allowed by physical laws. Our
results show that {\it the lower bound of the energy cost for the erasure of
information is zero}.  To this extent our results provide fresh insight into
the physical nature of information.

\begin{acknowledgements}
We are grateful to Viv Kendon and Martin Plenio for encouraging comments and
suggestions. SMB thanks the Royal Society and the Wolfson Foundation for
financial support and JAV acknowledges financial support from the Australian
Research Council.
\end{acknowledgements}

\end{document}